\documentclass[preprint,12pt]{elsarticle}

\usepackage{amssymb}

\usepackage{xcolor}
\usepackage[table]{xcolor}
\usepackage{amsmath}

\begin{document}

\begin{frontmatter}



\author{Julia Kończal\corref{cor1}}
\author{Michał Balcerek}
\author{Krzysztof Burnecki}
\address{Faculty of Pure and Applied Mathematics, Hugo Steinhaus Center,Wrocław University of Science and Technology,
            Wyb.
Wyspiańskiego 27, 
  50-370, 
            Wrocław,
            Poland}

\cortext[cor1]{Corresponding author. E-mail adress: julia.konczal@pwr.edu.pl}

\title{Machine learning models for predicting catastrophe bond coupons using climate data}


\begin{abstract}
 In recent years, the growing frequency and severity of natural disasters have increased the need for effective tools to manage catastrophe risk. Catastrophe (CAT) bonds allow the transfer of part of this risk to investors, offering an alternative to traditional reinsurance. This paper examines the role of climate variability in CAT bond pricing and evaluates the predictive performance of various machine learning models in forecasting CAT bond coupons. We combine features typically used in the literature with a new set of climate indicators, including Oceanic Niño Index, Arctic Oscillation, North Atlantic Oscillation,  Outgoing Longwave Radiation, Pacific–North American pattern, Pacific Decadal Oscillation, Southern Oscillation Index, and sea surface temperatures. We compare the performance of linear regression with several machine learning algorithms, such as random forest, gradient boosting, extremely randomized trees, and extreme gradient boosting. Our results show that including climate-related variables improves predictive accuracy across all models, with extremely randomized trees achieving the lowest root mean squared error (RMSE). These findings suggest that large-scale climate variability has a measurable influence on CAT bond pricing and that machine learning methods can effectively capture these complex relationships.
\end{abstract}




\end{frontmatter}
















\section{Introduction}

In recent years, natural disasters have become more frequent and severe around the world, leading to significant economic losses and growing financial risks. As a result, markets have started to pay greater attention to instruments that help transfer and mitigate catastrophe-related risk. One important example is catastrophe (CAT) bonds, which allow insurers to shift part of the potential losses from extreme weather events to investors. These bonds serve as an alternative to traditional reinsurance.

A CAT bond is typically issued by an insurer or reinsurer through a special purpose vehicle (SPV) \cite{braun2016pricing}. Investors who purchase the bond receive attractive coupon payments as long as no predefined catastrophic event (e.g., hurricane, earthquake, flood) occurs during the lifetime of the bond. However, if such an event takes place, part or all of the principal is used to cover the issuer’s losses, and investors can lose the corresponding portion of their investment. This mechanism aligns the interests of insurers seeking protection against extreme risks with investors willing to take on higher risk in exchange for higher potential returns.

Given the frequency and severity of climate-related catastrophes, accurate pricing of CAT bond coupons has become a critical challenge. Machine learning methods, which can capture nonlinear dependencies and extract patterns from large, multidimensional datasets, offer a promising alternative for pricing CAT bonds in a way that reflects evolving climate risks.

There is a vast literature on CAT bond pricing techniques. Early empirical research showed that expected loss is the main determinant of spreads at issuance, while factors such as covered territory, the identity of the sponsor, the reinsurance cycle, and corporate bond spreads also play important roles~\cite{braun2016pricing}. More recent work demonstrated that issuer effects strongly influence both CAT bond pricing and volatility, with implications for how different bond features behave across market phases and over time \cite{chatoro2023catastrophe}. Evidence on global warming and its implications for natural catastrophe risk further suggests that climate dynamics may contribute to a systematic undervaluation of climate-related risks in CAT bond markets \cite{morana2019climate}.

Another strand of the literature has focused on modeling catastrophe losses and developing approximations for tail probabilities to improve pricing methodologies. Weak approximations for heavy-tailed loss processes have been proposed and applied to index-linked CAT bond models \cite{burnecki2017stable}. Subsequent studies introduced pricing frameworks for multi-peril CAT bonds, modeling losses through compound non-homogeneous Poisson processes and fitting Burr, generalized extreme value, and log-normal distributions \cite{burnecki2024pricing}. Related extensions addressed heavy-tailed and left-truncated data, demonstrating advantages of alternative estimation techniques and their relevance for CAT bond pricing \cite{giuricich2019modelling}.
Contingent-claim approaches represent another important research direction. Models based on two-dimensional semi-Markov processes with stochastic interest rates have been proposed, yielding closed-form formulas calibrated to PCS data \cite{shao2017pricing}. Similar frameworks assumed that catastrophe losses follow compound non-homogeneous Poisson processes and relied on numerical approximations to obtain CAT bond prices \cite{ma2013pricing}.

In recent years, machine learning techniques have gained prominence in CAT bond pricing. Random forest models have been shown to significantly improve forecasts of CAT bond premia in both primary and secondary markets compared to traditional regression approaches \cite{gotze2020improving, gotze2023forecasting, makariou2021random}. Neural networks have also been applied to the design and pricing of earthquake-related CAT bonds, highlighting the broader potential of machine learning approaches for catastrophe-linked securities \cite{louloudis2024earthquake}.

Beyond CAT-bond–specific studies, related research in climate finance and catastrophe-linked derivatives offers insights relevant for pricing and risk assessment. Recent evidence shows that climate-change dynamics and natural disasters have measurable, heterogeneous effects on global financial markets, with climatological and biological events generating the strongest stock-market reactions and European markets displaying the highest sensitivity \cite{pagnottoni2022climate}. Furthermore, catastrophe-linked contingent claims have been examined through more flexible derivative-pricing approaches. A discrete-time pricing model for catastrophe equity put options introduces autocorrelated and catastrophe-dependent event intensity alongside Generalized Autoregressive Conditional Heteroskedasticity (GARCH) type stochastic volatility of the underlying asset, yielding an analytical pricing formula and demonstrating an inverted-U relationship between option prices, baseline catastrophe intensity, and the speed at which past events lose influence \cite{wang2019catastrophe}.

While previous studies have examined financial, structural, and climate-related determinants of CAT bond spreads, our analysis extends this literature in two key ways. First, we investigate whether large-scale climate variability indices—such as the Oceanic Niño Index (ONI), Arctic Oscillation (AO), North Atlantic Oscillation (NAO), Outgoing Longwave Radiation (OLR), the Pacific–North American (PNA) pattern, Pacific Decadal Oscillation (PDO), and the Southern Oscillation Index (SOI)—help explain CAT bond coupons, thereby enriching the set of climate variables considered in earlier work. Second, we evaluate the predictive performance of a broader range of machine learning models beyond random forest, including Bayesian ridge regression (BR), gradient boosting regression (GB), extremely randomized trees (ETR), automatic relevance determination regression (ARD), light gradient boosting machine (LGBM), and extreme gradient boosting (XGBoost), and compare their performance to traditional approaches. Overall, our study provides new insights into how climate variability may shape pricing in the CAT bond market and assesses the value added by alternative modeling techniques.

This paper is structured as follows. In Section \ref{sec:data} we explain the data under investigation, namely CAT bonds from the primary market. Section \ref{sec:climate} introduces the climate variability variables, including ONI, AO, NAO, PDO, and OLR, as well as sea surface temperatures (SST). We explore the correlations between lagged indices and CAT bond coupons. Section \ref{sec:LR} is devoted to comparing two models: the one proposed in \cite{braun2016pricing} and our extended model incorporating climate variability variables. We perform both point and interval predictions. In Section \ref{sec:ML}, we apply seven machine learning models (RF, BR, GB, ETR, ARD, LGBM, and XGBoost) to compare the performance of thebenchmark model and ours. Finally, Section \ref{sec:conclusions} concludes the paper.

\section{Data}
\label{sec:data}
\begin{figure}[h]
			\centering
			\includegraphics[width=1\linewidth]{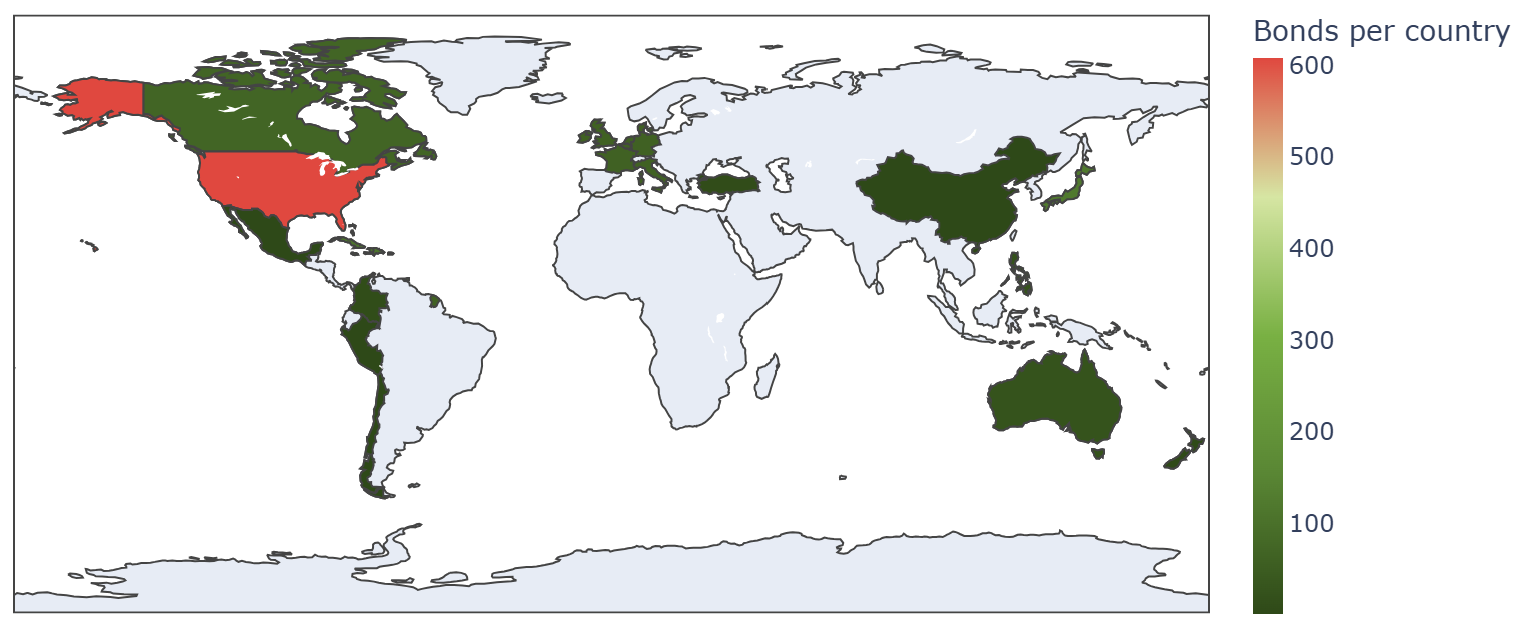}  
			\caption{Geographical distribution of bond issuance across countries. The  map presents the number of tranches issued in the primary market between June 1997 and December 2020.  }
			\label{fig:mapa}
	\end{figure}

Our dataset is collected from the primary market and consists of 734 tranches issued between June 1997 and December 2020. In Figure \ref{fig:mapa} we present geographical distribution of bond issuance across countries. As we can see the dataset covers a wide range of geographic regions. It captures both developed and emerging markets. The majority of bonds in the dataset provide coverage against catastrophe risks in the United States. However, we also observe issuances linked to exposures in Europe and Asia.

Here, we distinguish three groups of explanatory variables. First, we include CAT bond-related variables obtained directly from market data (e.g., expected loss, peril type, region). Second, we incorporate macroeconomic variables introduced in \cite{braun2016pricing}, while also including additional financial market indicators (monthly returns of major equity indices and selected cryptocurrencies). Third, we incorporate a set of climate variability indices.

\begin{table}[h]
\caption{Summary statistics for continuous variables in the dataset. The sample comes from the primary market and contains 734 tranches. All monetary values are expressed in millions of USD.}
\centering
\begin{tabular}{lc c c c}
\hline
 & Mean & Std&Min&Max \\
\hline
Attachment point       & 2505.3 & 3565.48 & 17.5&20670 \\
Attachment probability &0.039  & 0.039&0.00021&0.23\\
BB spread & 0.035&0.014&0.015&0.11\\
Cedent tenure &60.99 &70.93&0&281 \\
Coverage limit & 3267.14& 4331.26&65&	25000\\
Expected loss &0.025   &0.025&0&0.15 \\
Final spread price &0.077& 0.051&0.0065&0.49\\
No of locations & 1.34 & 0.65&1&3\\
No of perils & 2.37 &1.92&1&8 \\
ROL index & 220.49  &40.23&151.8&293.8 \\
Size &  134.28&123.59&1.8&1500 \\
Term & 36.38 & 12.5&1&120\\
\hline
\end{tabular}

\label{tab:stat1}
\end{table}

In Table~\ref{tab:stat1} we present summary statistics for continuous variables in our dataset. All monetary values are expressed in millions of USD. In particular, the table includes the mean, standard deviation, minimum, and maximum. The dataset contains 734 observations.

The attachment point represents the threshold above which the CAT bond begins to absorb losses. It varies across tranches, with a mean of \$2505.3 million and a standard deviation of \$3565.5. 
Related to this, the attachment probability reflects the modelled  probability that the bond will be triggered. This value ranges from nearly 0 to 23\%, with a mean of 3.9\%.

Turning to  market indicators, the BB spread is the average spread of BB-rated corporate bonds at issuance, serving as a benchmark for market pricing conditions. In our sample, the BB spread averages 3.5\%, ranging from 1.5\% to 11\%. Issuer-specific characteristics also play an important role. Cedent tenure measures the number of months the cedent has participated in the CAT bond market prior to a given issuance. This variable proxies for issuer experience. It ranges from 0 to 281 months, with a mean of 61 months.

The coverage limit corresponds to the maximum payout that the bond provides in the event of a qualifying catastrophe. The mean coverage limit is \$3.267 billion. Expected loss reflects the actuarial loss expectation over the bond's term. The average expected loss is 2.5\%, with values ranging from 0 to 15\%. The final spread price is the risk premium paid to investors, expressed as a percentage of notional. This is the dependent variable in our empirical analysis. The mean spread is 7.7\%. Throughout the paper, the terms “coupon”, “spread”, and “coupon rate” are used interchangeably to denote the risk premium paid to investors over the reference rate.

We also include the Rate-on-Line (ROL) index, which is an external index that tracks average pricing in the reinsurance market. The mean value is 220.5, with a range from 151.8 to 293.8. Tranche size denotes the value of each issuance. The average is \$134.3 million. Finally, the term of each bond, measured in months, reflects the contract duration. The average term is 36.4 months, with a minimum of 1 and a maximum of 120.5

\begin{table}[h]
\caption{Summary statistics for binary variables. The dataset contains 734 observations. Variables in Peril type, Trigger type, and Territory type are mutually exclusive, while observations may belong to multiple categories in the Region×Peril group.}
\centering
\begin{tabular}{l c c}
\hline
\textbf{Variable} & \textbf{Obs.} & \textbf{Percentage} \\
\hline

\rowcolor{gray!10}
\multicolumn{3}{c}{\textbf{Peril type}} \\[2pt]
Multiperil      & 408 & $55.59\%$  \\
Storm           & 177 & $24.11\%$  \\
Earthquake      & 127 & $17.3\%$   \\
Other           & 22  & $3\%$      \\[4pt]

\rowcolor{gray!10}
\multicolumn{3}{c}{\textbf{Trigger type}} \\[2pt]
Indemnity      & 316 & $43.05\%$  \\
Other trigger  & 418 & $56.95\%$  \\[4pt]

\rowcolor{gray!10}
\multicolumn{3}{c}{\textbf{RegionxPeril combination}} \\[2pt]
USxWind         & 459 & $62.53\%$   \\
USxEarthquake   & 407 & $55.45\%$   \\
EuropexWind     & 138 & $18.8\%$    \\
JapanxEarthquake & 89 & $12.13\%$   \\

\rowcolor{gray!10}
\multicolumn{3}{c}{\textbf{Territory type}} \\[2pt]
Multiterritory & 188 & $25.61\%$  \\
US             & 420 & $57.22\%$  \\
Europe         & 54  & $7.36\%$   \\
Japan          & 45  & $6.13\%$   \\
Other territory& 31  & $4.22\%$   \\

\hline
\end{tabular}
\label{tab:stat2}
\end{table}

In Table~\ref{tab:stat2} we provide descriptive statistics for the binary variables. The dataset contains a total of 734 observations. The Percentage column reports the share of each category relative to all observations. Variables in the categories Peril type, Trigger type, and Territory type are mutually exclusive, whereas observations may fall into multiple categories within the Region×Peril combination group.

The Multiperil variable indicates whether a tranche covers more than one type of peril. This applies to 55.6\% of bonds. Among individual peril types, storm risks are present in 24.1\% of tranches, earthquake risks in 17.3\%, and other perils (e.g., flood, wildfire, pandemic) in 3\%.  With respect to trigger mechanisms, 43.1\% of bonds use an indemnity trigger, which pays out based on actual losses incurred by the issuer. The remaining 56.9\% employ non-indemnity triggers, such as industry loss, parametric, or modelled loss mechanisms.

The dataset also includes several region-peril variables, indicating whether the bond covers specific combinations such as U.S. wind (62.5\%), U.S. earthquake (55.5\%), Europe wind (18.8\%), and Japan earthquake (12.1\%). Multiterritory bonds (25.6\%) provide coverage across more than one geographic region, while single-region variables show that 57.2\% of tranches cover the U.S., 7.4\% Europe, 6.1\% Japan, and 4.2\% other territories (e.g., Australia, Mexico, Caribbean countries).

\section{Climate variability variables}
\label{sec:climate}

Climate variability indices capture large-scale, recurring patterns in ocean-atmosphere systems that influence the frequency and severity of extreme weather events worldwide. We consider 7 climate variability indicators as well as sea surface temperatures across various oceanic regions.

The first indicator we consider is the ONI, which captures SST anomalies in the central Pacific (Niño 3.4 region) \cite{glantz2020reviewing}. The ONI is the primary measure of the El Niño–Southern Oscillation (ENSO) phenomenon, where positive values indicate El Niño conditions and negative values indicate La Niña. 

Next, we include the AO, an index representing atmospheric pressure variability over the Arctic and mid-latitudes \cite{lorenz1951seasonal}. The AO describes the strength of the polar vortex. In its positive phase, it tends to confine colder air to the polar regions, whereas its negative phase allows cold air to spill into mid-latitudes, increasing the likelihood of extreme weather in North America and Europe.

The third indicator is the NAO, which reflects pressure differences between the Azores High and the Icelandic Low \cite{visbeck2003north}. The NAO significantly influences winter weather, storm tracks, and precipitation across the North Atlantic region, especially in Europe. 

We also consider OLR, a measure of the Earth’s radiative heat loss to space \cite{singh2022thermodynamics}. OLR serves as a proxy for tropical convection and cloud cover, with lower values generally indicating stronger convective activity and potential rainfall. 

The fifth indicator is the PNA pattern, a mode of atmospheric variability associated with pressure anomalies across the North Pacific and North America \cite{franzke2011synoptic}. The PNA affects the jet stream and plays a critical role in shaping winter climate conditions across the United States and Canada, with implications for windstorm and cold-weather risks.

Another index included in our analysis is the PDO, which tracks multi-decadal fluctuations in North Pacific sea surface temperatures \cite{deser2010sea}. Positive and negative phases of the PDO are linked to changes in regional climate conditions such as droughts, floods, and temperature anomalies. 

Finally, we examine the SOI, which measures sea-level pressure differences between Tahiti and Darwin \cite{ranasinghe2004southern}. Like the ONI, the SOI is a core ENSO indicator, with strongly negative values corresponding to El Niño episodes.

\begin{figure}[h]
			\centering
			\includegraphics[width=1\linewidth]{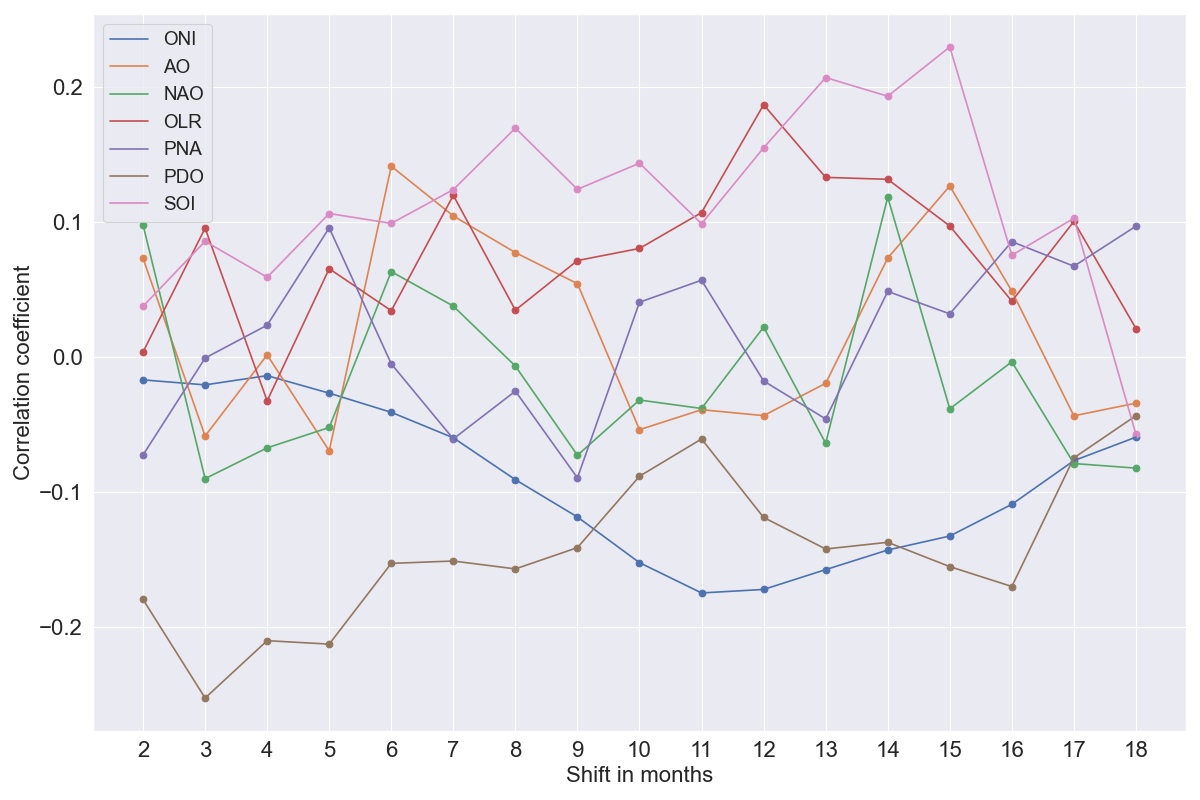}  
			\caption{Correlation between catastrophe bond coupon rates and lagged values of large-scale climate indices. The x-axis shows the number of months each climate index is shifted backward to capture delayed effects, while the y-axis presents the corresponding correlation coefficient.}
			\label{fig:corr}
	\end{figure}

 In Figure \ref{fig:corr}  we present the correlation coefficients between CAT bond coupon and lagged values of climate indices \cite{mantegna1999introduction}. Each index is shifted backward in time from 2 to 18 months to capture potential delayed effects. The ONI exhibits negative correlations with coupon values, with the lowest values between 9 and 14 months lag. This suggests that warmer ENSO phases might be associated with lower CAT bond coupons. The AO shows positive correlations across most lags, peaking slightly at around 3-5 months and again at 14-16 months. The NAO displays  positive correlations at shorter lags (around 3-5 months) and again at 12-13 months, with a peak correlation close to 0.12. The OLR stands out as one of the most consistently positively correlated indices, especially at higher lags between 11 and 16 months, where correlations approach or exceed 0.2. Since OLR is a proxy for tropical convection and rainfall, these findings may indicate that sustained tropical activity affects perceived risk in CAT bond markets. The PNA also shows a relatively stable and positive relationship with CAT bond coupons. Correlations range from 0.05 to over 0.2. 
In contrast, the PDO consistently exhibits negative correlations across all lags, with a particularly strong values at short lags between 2 and 4 months.
Finally, the SOI shows a broadly positive correlation values, especially between 5 and 15 months. Correlations peak above 0.2 around month 14. Since SOI and ONI are both ENSO-related indices, this pattern reinforces the observed link between ENSO dynamics and CAT bond pricing.

Overall, the figure reveals that OLR, SOI, and PNA exhibit the strongest and most consistent positive correlations with CAT bond coupons, particularly at lags above 10 months. This suggests that large-scale climate variability, especially those linked to tropical convection and North American atmospheric patterns, may have delayed yet measurable impacts on CAT bond pricing mechanisms. In contrast, the PDO and ONI indices are negatively or weakly correlated with spreads.

\begin{figure}[h]
			\centering
			\includegraphics[width=1\linewidth]{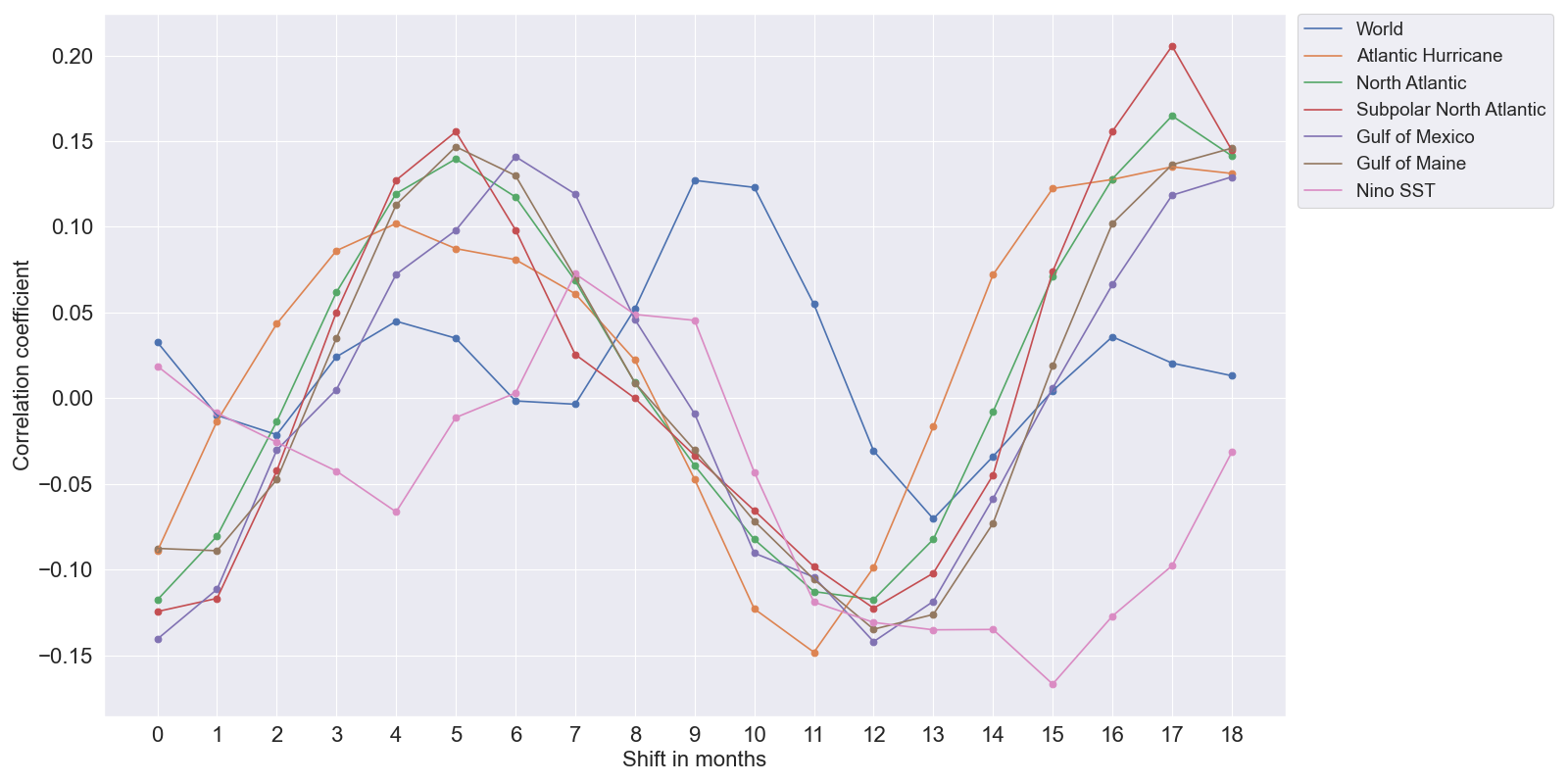}  
			\caption{Correlation between catastrophe bond coupon rates and lagged SST across various oceanic regions. The $x$--axis represents the number of months by which SST values are shifted backward, and the $y$--axis shows the corresponding correlation coefficient.}
			\label{fig:sst_mean_2020}
	\end{figure}

In Figure~\ref{fig:sst_mean_2020} we present the correlation coefficient between catastrophe bond coupon rates and lagged SST anomalies across seven oceanic regions: World (global mean SST), Atlantic Hurricane, North Atlantic, Subpolar North Atlantic, Gulf of Mexico, Gulf of Maine, and Niño SST (Niño 3.4 region). SST values were shifted backward by 0 to 18 months to assess whether delayed oceanic thermal patterns have explanatory power for CAT bond pricing.

Taken together, the figure suggests that regional SST anomalies, particularly in the Atlantic Hurricane region, Gulf of Mexico, and North Atlantic, are more strongly correlated with CAT bond coupons than global or Niño SST indicators. 

\section{Linear regression comparison}
\label{sec:LR}
We compare the explanatory power of two models. The first model replicates the one proposed in \cite{braun2016pricing}, serving as a benchmark. The second is our extended model, which incorporates additional variables related to climate variability.

The benchmark model follows the original approach and includes a comprehensive set of bond-specific and structural variables. These include catastrophe bond fundamentals such as expected loss, size, term, and trigger type, as well as peril and region-specific dummies (Wind, Earthquake, Multiterritory, US, Europe, Japan, and their interactions). It also controls for issuer quality (Swiss Re, investment grade), market conditions (ROL index, BB spread).

The extended model builds upon this structure by incorporating additional variables. In particular, it includes the ROL index change, which represents the annual change in the ROL index. We also include two lagged climate indices: SOI shifted by 15 months and OLR shifted by 12 months. These additions aim to test whether climate-related variables carry additional explanatory power beyond traditional CAT bond characteristics.

For the feature selection for the extended model we used the Elastic Net regularization method \cite{zou2005regularization}. This approach combines the $L_1$ penalty from Lasso regression with the $L_2$ penalty from Ridge regression. By balancing these two components, Elastic Net selects the most relevant predictors.
This procedure allowed us to identify the subset of climate and market variables that contributed most to explaining the variation in CAT bond coupons.
The dependent variable in both models is the final spread price, and all variables have been standardized or transformed as necessary for comparability.

The evaluation of point forecasts shows that both models perform well in terms of predictive accuracy. For the benchmark model, the OLS regression achieves mean squared error $MSE = 0.000315$, mean absolute error $MAE = 0.01323$, and root mean squared error $RMSE = 0.01774$ on the test sample. For the extended model, which incorporates climate-related variables, test errors improve to $MSE = 0.000283$, $MAE = 0.01279$, and $RMSE = 0.01682$. These results confirm that adding the climate indicators improves the explanatory and predictive power of the model.

In addition to point forecasts, we also conduct a probabilistic forecasting exercise to evaluate the uncertainty surrounding the predicted catastrophe bond coupons. The dataset is divided into three non-overlapping subsets in an 80:10:10 ratio. The first part is used to estimate model parameters, the second serves as a calibration window, and the third is reserved for out-of-sample testing. In the calibration window, we generate point forecasts, compute residuals, and fit a normal distribution to the forecast errors. Based on these residuals, we also calculate empirical quantiles to capture the historical distribution of forecast uncertainty. In the final test window, we produce probabilistic forecasts by adding 1,000 random draws from the fitted normal distribution to each model’s point prediction. This approach yields full predictive distributions for each observation, from which we derive probabilistic measures such as the 5th, 50th, and 95th percentiles.

To assess the quality of the probabilistic forecasts, we apply the Kupiec and Christoffersen tests for Value-at-Risk (VaR) at the 5\% level, including tests for unconditional coverage (LRUC), independence (LRIND), and conditional coverage (LRCC), along with the Basel traffic-light classification \cite{christoffersen1998evaluating, kupiec1995techniques}. For the benchmark model, the percentage of failures equals 4.35\%, with $LRUC = 0.0645 $ $(p = 0.7995)$, $LRIND = 0.277$ $(p = 0.5987)$, and $LRCC = 0.3415$ $(p = 0.843)$, placing the model in the green Basel zone. For the extended model, the percentage of failures equals 1.45\%, with $LRUC = 2.5137$ $(p = 0.1129)$, $LRIND = 0.0299$ $(p = 0.8628)$, and $LRCC = 2.5436$ $(p = 0.2803)$, also classified as green. These outcomes suggest that both models generate well-calibrated probabilistic forecasts, with the extended model producing fewer exceedances in the lower tail. Overall, the results show that extending the model with climate-related variables improves not only point forecast accuracy but also the stability and reliability of probabilistic predictions.

\section{Machine learning approach}
\label{sec:ML}

We applied linear regression, which allowed us to directly compare the benchmark feature set with our proposed feature set \cite{braun2016pricing}. In the next step, we extend this comparison by applying various machine learning models using both benchmark feature set and our own, to evaluate how different algorithms perform across the two models. We apply 7 models: Random Forest, Bayesian Ridge Regression, Gradient Boosting Regression,  Extremely Randomized Trees, Automatic Relevance Determination Regression, Light Gradient Boosting Machine, and Extreme Gradient Boosting.
All models are trained on the same dataset 
\(\mathcal{D} = \{(x_i, y_i)\}_{i=1}^N\), 
where \(x_i \in \mathbb{R}^p\) denotes the vector of features and \(y_i \in \mathbb{R}\) the target variable (CAT bond coupon).  
The goal is to learn a function \(f: \mathbb{R}^p \rightarrow \mathbb{R}\) that minimizes prediction error on unseen data.

Random Forest is an ensemble method that combines multiple decision trees \cite{james2013introduction}.  
Each tree is trained on a bootstrap sample of the dataset, 
and at each split only a random subset of features is considered, 
which helps to reduce correlation between trees.  
Formally, the algorithm builds $T$ regression trees $\{h_t(x)\}_{t=1}^T$, 
and the final prediction is obtained by averaging their outputs:
\begin{equation}
   \hat{y}(x) = \frac{1}{T} \sum_{t=1}^T h_t(x).
\end{equation}


Bayesian Ridge Regression is a linear regression model that introduces a probabilistic 
framework by placing prior distributions over the model parameters \cite{tipping2001sparse}.
This approach allows automatic regularization and provides posterior distributions 
instead of point estimates.  
The regression model is defined as
\begin{equation}
 y_i = x_i^\top w + \varepsilon_i, \quad \varepsilon_i \sim \mathcal{N}(0, \alpha^{-1}),   
\end{equation}
where $w \in \mathbb{R}^p$ is the weight vector and $\alpha$ denotes the noise precision. 
During hyperparameter tuning, $\alpha$ was searched over the range 
$\alpha \in [10^{-6}, 10^{3}]$.

A zero-mean Gaussian prior is placed over the coefficients:
\begin{equation}
  p(w \mid \lambda) \sim \mathcal{N}(0, \lambda^{-1} I_p),  
\end{equation}
where $\lambda$ controls the strength of regularization.
The prior precision parameter was tuned over 
$\lambda \in [10^{-6}, 10^{3}]$.

Gradient Boosting is an ensemble method that builds a strong predictor by combining 
multiple weak learners (typically regression trees) in a stage-wise manner \cite{friedman2003multiple}.
Each new tree is trained to minimize the residual errors of the previous ensemble 
using gradient descent in function space.  
The model is expressed as an additive expansion
\begin{equation}
F_m(x) = \sum_{k=1}^m \gamma_k h_k(x),
\end{equation}
where $h_k(x)$ denotes the $k$-th regression tree (weak learner) and $\gamma_k$ is its weight.  
The method optimizes a differentiable loss function $L(y, F(x))$ by iteratively fitting new trees 
to the negative gradient of the loss with respect to the current model predictions.  
At iteration $m$, the pseudo-residuals are computed as
\begin{equation}
r_{im} = - \left[ \frac{\partial L(y_i, F(x_i))}{\partial F(x_i)} \right]_{F(x) = F_{m-1}(x)}.
\end{equation}
for $i = 1,2, \ldots, N$.
A new regression tree $h_m(x)$ is fitted to the residuals $\{r_{im}\}$.  
The optimal step size $\gamma_m$ is obtained by solving
\begin{equation}
\gamma_m = \mathrm{argmin}_\gamma \sum_{i=1}^N L\big(y_i, F_{m-1}(x_i) + \gamma h_m(x_i)\big).
\end{equation}
The model is then updated as
\begin{equation}
F_m(x) = F_{m-1}(x) + \nu \gamma_m h_m(x),
\end{equation}
where $\nu \in (0,1]$ is the learning rate controlling the contribution of each tree.

The Extremely Randomized Trees algorithm is an ensemble method based on aggregating 
a large number of randomized decision trees \cite{geurts2006extremely}.
It is related to RF but introduces additional randomization in the tree-building process.  
The algorithm builds $T$ regression trees $\{h_t(x)\}_{t=1}^T$.  
Each tree is trained on the entire dataset (unlike RF, where bootstrap samples are used).  
At each split of a tree, a random subset of features \( M \subset \{1, \dots, p\} \) is first selected. 
For every feature in this subset, random split thresholds are then drawn uniformly from the observed range of that feature. 
Among all possible splits, the algorithm chooses the one that maximizes the reduction in variance, 
following a standard variance reduction criterion:

\begin{equation}
\Delta(S, f, \theta) = \mathrm{Var}(S) - \frac{|S_L|}{|S|}\mathrm{Var}(S_L) - \frac{|S_R|}{|S|}\mathrm{Var}(S_R),
\end{equation}
where $S$ is the current sample set at a node, $f$ is the feature, 
$\theta$ is the split threshold, and $S_L, S_R$ are the left and right subsets after splitting. 

Automatic Relevance Determination regression is a Bayesian linear model 
that extends ridge regression by placing independent priors on each coefficient \cite{li2002bayesian}. 
The model is expressed as
\begin{equation}
y_i = x_i^\top w + \varepsilon_i, \quad \varepsilon_i \sim \mathcal{N}(0, \alpha^{-1}),
\end{equation}
where $w \in \mathbb{R}^p$ is the weight vector and $\alpha$ is the precision 

In ARD, each coefficient $w_j$ has an individual Gaussian prior:
\begin{equation}
p(w \mid \{\lambda_j\}) \sim \prod_{j=1}^p \mathcal{N}(w_j \mid 0, \lambda_j^{-1}),
\end{equation}
where the precision parameters were tuned jointly over the range 
$\lambda_j \in [10^{-6}, 10^{3}]$.

Noise precision was included in tuning as 
$\alpha \in [10^{-6}, 10^{3}]$.
Applying Bayes’ theorem, the posterior distribution of the weights is
\begin{equation}
p(w \mid X, y, \alpha, \{\lambda_j\}) \sim \mathcal{N}(w \mid \mu_w, \Sigma_w),
\end{equation}
with
\begin{equation}
\Sigma_w = \left(\mathrm{diag}(\lambda_1, \dots, \lambda_p) + \alpha X^\top X \right)^{-1}, 
\end{equation}
\begin{equation}
\mu_w = \alpha \Sigma_w X^\top y,
\end{equation}
where $X \in \mathbb{R}^{N \times p}$ is the design matrix.

Light Gradient Boosting Machine is a gradient boosting framework 
based on decision trees \cite{kopitar2020early}.
It improves upon standard gradient boosting by using optimized techniques such as 
histogram-based splitting and leaf-wise tree growth.  
LGBM constructs an additive model of regression trees:
\begin{equation}
F_m(x) = \sum_{k=1}^m \gamma_k h_k(x),
\end{equation}
where $h_k(x)$ is the $k$-th regression tree and $\gamma_k$ is its weight.  
At iteration $m$, the objective is to minimize a differentiable loss function $L(y, F(x))$.  
Using a second-order Taylor expansion, the loss is approximated as
\begin{equation}
L^{(m)} \approx \sum_{i=1}^N \left[ g_{im} h_m(x_i) + \tfrac{1}{2} h_{im} h_m(x_i)^2 \right],
\end{equation}
where
\begin{equation}
g_{im} = \frac{\partial L(y_i, F(x_i))}{\partial F(x_i)}, 
\quad 
h_{im} = \frac{\partial^2 L(y_i, F(x_i))}{\partial F(x_i)^2}
\end{equation}
are the first and second derivatives of the loss with respect to predictions.  
The optimal leaf values of the tree are computed using both gradients $g_{im}$ 
and Hessians $h_{im}$, which improves convergence compared to first-order methods.  
The model is updated iteratively as
\begin{equation}
F_m(x) = F_{m-1}(x) + \nu \gamma_m h_m(x),
\end{equation}
where $\nu \in (0,1]$ is the learning rate controlling the contribution of each tree.  

Extreme Gradient Boosting is an efficient and scalable implementation 
of gradient boosting that incorporates additional regularization to improve 
generalization and prevent overfitting \cite{carmona2019predicting}.  
It has become one of the most widely used boosting algorithms for both regression 
and classification tasks.  
XGBoost constructs an additive model of regression trees:
\begin{equation}
F_m(x) = \sum_{k=1}^m h_k(x),
\end{equation}
where $h_k(x)$ denotes the $k$-th regression tree.  
The training objective at iteration $m$ is defined as
\begin{equation}
\mathcal{L}^{(m)} = \sum_{i=1}^N L(y_i, F_{m-1}(x_i) + h_m(x_i)) + \Omega(h_m),
\end{equation}
where $L(y, \hat{y})$ is a differentiable convex loss function, 
and $\Omega(h_m)$ is a regularization term that penalizes model complexity.  
Using a second-order Taylor expansion, the objective can be approximated as
\begin{equation}
\mathcal{L}^{(m)} \approx \sum_{i=1}^N \left[ g_{im} h_m(x_i) + \tfrac{1}{2} h_{im} h_m(x_i)^2 \right] + \Omega(h_m),
\end{equation}
where
\begin{equation}
g_{im} = \frac{\partial L(y_i, F(x_i))}{\partial F(x_i)}, 
\quad 
h_{im} = \frac{\partial^2 L(y_i, F(x_i))}{\partial F(x_i)^2}.
\end{equation}
The regularization term is defined as
\begin{equation}
\Omega(h_m) = \gamma T + \frac{1}{2} \lambda \sum_{j=1}^T w_j^2,
\end{equation}
where $T$ is the number of leaves in the tree, $w_j$ is the weight of leaf $j$, 
$\gamma$ penalizes the number of leaves, and $\lambda$ controls $L_2$ regularization 
on leaf weights.  
The optimal weight for each leaf is given by
\begin{equation}
w_j^* = - \frac{\sum_{i \in I_j} g_{im}}{\sum_{i \in I_j} h_{im} + \lambda},
\end{equation}
where $I_j$ is the set of samples assigned to leaf $j$.  
The corresponding optimal value of the objective function is
\begin{equation}
\mathcal{L}^{(m)}_{\text{opt}} = -\frac{1}{2} \sum_{j=1}^T \frac{\left( \sum_{i \in I_j} g_{im} \right)^2}{\sum_{i \in I_j} h_{im} + \lambda} + \gamma T.
\end{equation}

To ensure a fair comparison across methods, we performed hyperparameter tuning for all machine learning models. We applied a randomized search combined with 5-fold cross-validation, using the training set only. For each model, a predefined search space of hyperparameters was explored, and the configuration minimizing the cross-validated RMSE was selected. Tree-based models (RF, ETR, GB, LGBM, XGBoost) were tuned over parameters controlling tree depth, number of estimators, subsampling, and regularization strength, while linear Bayesian models (BRR and ARD) were tuned over their respective prior precision parameters. The final results reported below are based on models refitted on the full training data using the best-performing hyperparameter sets.

\begin{table}[ht]
\caption{Values of the RMSE calculated for the predictions on the testing set, rounded to five significant figures.}
\centering
\begin{tabular}{lc c}
\hline
Model & Benchmark model & Our model \\
\hline
OLS       & 0.018443 & 0.016823 \\
RF & 0.018844 & 0.017563 \\
BRR & 0.018447 & 0.016813 \\
GBR & 0.019155 & 0.017756 \\
ETR & 0.014161  & $\mathbf{0.012294}$ \\

ARD & 0.017709 & 0.017085 \\
LGBM & 0.019088 & 0.018858 \\
XGB & 0.018487 & 0.018163 \\
\hline
\end{tabular}

\label{tab:RMSE_test}
\end{table}

In Table \ref{tab:RMSE_test} we present the RMSE on the test set 
for both benchmark feature set and our proposed features, across all evaluated models. These results correspond to point forecasts, where each model provides a single predicted value 
of the CAT bond coupon for each observation. 
 The baseline model (OLS) shows slightly lower error with our features (0.0168) 
compared to benchmark (0.0184). A similar pattern can be observed for all models, 
indicating that our representation generally improves predictive accuracy.  
Among tree-based ensembles, the ETR algorithm achieves 
the lowest overall error, with RMSE of 0.0123 for our features, which is 
a notable improvement over benchmark 0.0142 (approximately 13.2\%). 
Other boosting methods such as Gradient Boosting, LGBM, and XGBoost 
show consistent but smaller gains.  
Linear Bayesian models (BRR and ARD) also benefit from our features, though 
the improvement is less pronounced. Overall, the comparison demonstrates that 
our feature set leads to systematically better results across a range of 
algorithms, with the largest relative gain observed for ETR. We also explored probabilistic forecasts for the machine learning models using the same procedure as in the linear regression framework, and the results were consistent with those reported in Section~\ref{sec:LR}, confirming that the inclusion of climate-related variables improved both point and probabilistic predictive performance across all approaches.

\section{Conclusions}
\label{sec:conclusions}

CAT bonds are insurance-linked securities designed to transfer catastrophe risk from insurers and reinsurers to capital market investors. Introduced in the 1990s following a series of major natural disasters, these instruments allow issuers to obtain protection against extreme losses by issuing debt that is partially or fully forgiven in the event of a qualifying catastrophe. In return, investors receive attractive coupon payments as compensation for taking on this risk. The CAT bond market has grown over the past two decades, reaching tens of billions of dollars in outstanding notional value, and has become a crucial component of global risk management and alternative reinsurance. Because CAT bonds directly link financial performance to natural hazard outcomes, their pricing reflects not only traditional financial factors but also evolving climate and environmental risks.

In this study, we investigated whether climate variability helps explain and predict CAT bond coupons. We combined standard bond- and market-related features used in the literature with climate indicators such as ONI, SOI, AO, NAO, PDO, PNA, OLR, and regional SST. Our dataset covers 734 primary-market tranches issued between June 1997 and December 2020.

Before conducting predictive modeling, we performed an extensive correlation analysis between CAT bond coupon rates and lagged values of key climate indices. The results revealed several patterns. OLR, SOI, and PNA exhibited the strongest positive correlations with CAT bond coupons, especially at lags of 10–16 months, suggesting that tropical convection and North American atmospheric variability have delayed yet measurable impacts on market pricing. In contrast, the PDO and ONI indices showed persistently negative or weak correlations, while regional SST anomalies, particularly in the Atlantic Hurricane region, Gulf of Mexico, and North Atlantic, were more strongly associated with coupon variation than global or Niño SST averages.

Building upon these insights, we compared two model specifications: a benchmark that replicates the model proposed in \cite{braun2016pricing} and an extended feature set that extends the feature set by climate-related variables. Using an 80:10:10 split, the first window was used for estimation, the second for calibration, and the third for out-of-sample testing. For point forecasts, ordinary least squares (OLS) with the benchmark features achieved a train $R^2$ of 0.833 and test errors of MSE $=0.000315$, MAE $=0.01323$, and RMSE $=0.01774$. Our specification improved performance to a train $R^2$ of 0.835 and reduced test errors to MSE $=0.000283$, MAE $=0.01279$, and RMSE $=0.01682$. These results indicate that adding climate information improves point prediction accuracy.

Next, we compared several machine learning algorithms. Across models, the extended feature set lowered test RMSE compared to the benchmark model. Among tree-based ensembles, Extremely Randomized Trees achieved the best results, with RMSE $=0.0123$ for the extended features compared with $0.0142$ for the benchmark features. Gradient Boosting, LightGBM, XGBoost, and Bayesian linear models (BRR, ARD) also showed smaller error values when climate variables were included. Overall, these findings confirm that flexible, nonlinear learners can effectively increase predictive accuracy.

Finally, we constructed probabilistic forecasts to quantify uncertainty around coupon predictions. Residuals from the calibration window were used to fit a normal error distribution and to compute empirical quantiles. In the test window, we generated predictive distributions by adding 1{,}000 Monte Carlo draws from the fitted error law to each point forecast and then evaluated 5\% Value-at-Risk coverage. For the benchmark specification, the percentage of failures was 4.35\%, with $LRUC =0.0645$ $(p=0.7995)$, $LRIND =0.277$ $(p=0.5987)$, and $LRCC =0.3415$ $(p=0.843)$, yielding a Basel green classification. For the extended model, percentage of failures was 1.45\%, with $LRUC =2.5137$ $(p=0.1129)$, $LRIND =0.0299$ $(p=0.8628)$, and $LRCC =2.5436$ $(p=0.2803)$, also in the green zone. These results show that both models produce well-calibrated tails.

Taken together, the evidence shows that climate variability provides additional important information for CAT bond pricing and that machine learning methods can capture this information to improve both point and probabilistic forecasts. While no single model dominates every metric, the Extremely Randomized Trees seems to perform the best, making it a practical choice for forecasting CAT bond coupons.

\section*{Acknowledgements}
We would like to express our gratitude to Alexander Braun for providing access to the dataset used in this study. The work of JK and KB was supported by NCN Grant No. 2022/47/B/HS4/02139.

\bibliographystyle{elsarticle-num} 
\bibliography{references}



\end{document}